\documentclass[onecolumn,noshowpacs,preprintnumbers,amsmath,amssymb]{revtex4}  

\usepackage{graphicx}
\usepackage{dcolumn}
\usepackage{bm}
\usepackage{subfigure}
\usepackage{color}
\def\be{\begin{equation}}
\def\ee{\end{equation}}
\def\bee{\begin{enumerate}}
\def\eee{\end{enumerate}}

\begin{document}

\title{How aggressive a driver is? - A quantitative analysis\\
} 

\author{Subinay Dasgupta}
\affiliation{Department of Physics, University of Calcutta,
92 Acharya Prafulla Chandra Road, Kolkata 700009, India}

\author{Sitabhra Sinha}
\affiliation{The Institute of Mathematical Sciences, CIT Campus - Taramani, Chennai 600113, India}

\begin{abstract}
{\bf Consider a bottleneck in a road through which only one car can pass through. Suppose that at a time the car passing will have the most aggressive driver in queue and that the aggressiveness of an individual is measured by an attribute $A \equiv N \tau^{\sigma}$ where the quantity $N$  varies randomly from person to person in the range 0 to 1, $\tau$ is the time for which the driver is waiting in the bottleneck and the parameter $\sigma$ is the same for all individuals. Thus, we assume that the aggressiveness depends on the nature of the individual and increases with waiting time in a traffic jam. In support of the algebraic form of $A$, we show (numerically and analytically) that our hypothesis implies that the probability of waiting for a time $\tau$ will be $P(\tau) \propto \tau^{\alpha}$ with the value of $\alpha$ fixed by $\sigma$. Empirical studies confirm such variation in $P(\tau)$ with an exponent of 3.0 to 3.5 in two different cities of India and 1.5 for a traffic intersection in Germany. There is a possibility that the parameter $\sigma$ (and hence $\alpha$) is characteristic of a geographical region.}
\end{abstract}

\maketitle

{\bf Introduction:} Many features of human civilisation originate from imperfections of human behaviour. One example of this is the effect of limited rationality and limited self-control on market economy, as observed by Richard Thaler \cite{Thaler}.  In this article, we shall analyse quantitatively the effect of {\em aggressiveness} in car-driving on the pattern of traffic jam. The importance of aggressiveness of driver on traffic movement has been emphasized earlier \cite{agr-prev} from various perspectives. In general, this parameter will vary from person to person and for a given person depend on various factors like psychological condition, density of traffic around and is likely to increase as one waits longer in a congestion.  We start with the hypothesis that the aggressiveness of a driver at a moment in a given surroundings can be quantified by a parameter $A$ 
and depends only on the intrinsic nature of the person and the duration for which he/she has been stuck up in traffic congestion. Our second hypothesis is that the quantity $A$  is  the arithmetic product of an attribute $N$ (with numerical value between 0 and 1) specific to the nature of an individual and a function $f(\tau)$ of the congestion time $\tau$. (We measure $\tau$ as the time for which the speed of the vehicle has been less than some threshold value.) The quantity $f(\tau)$ increases with $\tau$ and is the same for all individuals and we present justifications for the functional form $f(\tau) = \tau^{\sigma}$  where $\sigma$ is  a measure of how rapidly the arrogance of a driver increases with increase of waiting time. 

In order to test the validity of our hypotheses, 
we consider a physically measurable quantity, namely  the probability $P(\tau)$ of being stranded for time $\tau$. Following a reasonable algorithm, we calculate $P(\tau)$ analytically and by simulation and find that it decays algebraically with $\tau$ 
\be P(\tau) \sim \frac{1}{\tau^{\alpha}} \label{Ptau} \ee
(with the value of $\alpha$ being fixed by that of $\sigma$). The crucial point is that such behaviour agrees with the observations on traffic congestion in India and Germany (see below). We claim that such algebraic decay and the specific values of the exponent $\alpha$ arise from a psychological imperfection namely {\em aggressiveness} of the car drivers. Appearance of algebraic decay in the context of human behaviour has been reported earlier also \cite{Zhao}\\

{\bf Data Analysis:} Empirical data for the time interval for which vehicles are waiting in a traffic jam, were collected from GPS data of about 1000 radio-taxis in 2 cities in India \cite{IMSc} and it was found that the distribution of waiting time obeys Eq. (\ref{Ptau}) with $\alpha \sim 2.5 $ to 3.5 for Bangalore and Delhi.   A similar study for a traffic cross section in Germany \cite{Krause} gives a value $\alpha = 1.5$. \\

{\bf The Model:} 
We shall develop a cellular automata (a rule for movement of vehicles) following the hypotheses mentioned above. Suppose we have a road narrowed down at a place in such a way that only one car can pass through. For a large number of incoming cars, all are queued up before the  spot and they pass one by one. At one given time-step the car that passes through has the most {\em aggressive} driver, that is, the one whose driver has the largest value of $A$ defined according to our hypotheses as
\be A = N f(\tau) = N \tau^{\sigma} \label{defA} \ee
The precise algorithm adopted is as follows: We start at time $t=0$ with a queue of $L$ cars having parameters $N_1, N_2, \cdots N_L$ drawn from a uniform distribution in the range 0 to 1. At each update $t\to t+1$, the car with highest $A$-value goes off and is replaced by another car. (If multiple cars have the highest $A$-value, then any one of them, chosen randomly, goes off.)  The waiting time for the removed car is $\tau = t-t_0$, where $t_0$ is the time-step at which it was introduced. The new car has an $N$ value $ \in (0,1)$ and the value of $t_0$ for this new member is $t$. The number of cars queued up remains the same with time. This model is a variant of the queuing model of Barabasi et.al. \cite{B1, B2}.\\
 
Numerical simulation shows that the probability distribution for the waiting time decays algebraically as $1/\tau^{\alpha}$, where the exponent $\alpha$ depends on $\sigma$ and, weakly, on the number of cars $L$. Typically, for $\sigma=0.5$, the exponent $\alpha$ varies from 1.77 to 1.53 as $L$ varies from 10 to 100. For $\sigma=2.0$, the value of $\alpha$ ranges from 3.2 to 3.0 for a similar variation of $L$ (see Fig. 1(a, b)). We shall show analytically that $\alpha = \sigma +1$ when fluctuations are ignored. This shows that as the aggressiveness of the drivers grows more rapidly with waiting, the index $\alpha$ increases. When $\sigma$ is zero, that is, the aggressiveness of the driver does not increase with waiting, the index $\alpha$ is 1.0 as has been observed in the queuing model of Barabasi et.al. \cite{B1, B2}.

In support of the functional form $f(\tau) = \tau^{\sigma}$, we observe that (i) this functional form does reproduce the behaviour of Eq. (\ref{Ptau}) (ii) other functional forms (for example, $f(\tau) = \exp(\sigma \tau)$ and the assumption that $\sigma$ varies from person to person following a uniform random distribution  $ \in (0,1)$) does not reproduce Eq. (\ref{Ptau}). 

Till now we have worked with the rather unrealistic assumption that the number of cars queued up does not vary with time. We have checked that when the number of cars vary randomly  about a mean value, the results remain qualitatively the same.
We have also studied a variant of our model, where the front row just before the bottleneck can accommodate $w$ cars only and multiple rows (each with $w$ cars) are formed. The algorithm for updating $t \to (t+1)$ is : (i) look at the front row and remove the member with largest $A$; (ii) the waiting time for the removed car is $t-t_0$, where $t_0$ is the time-step at which it was introduced; (iii) replace the removed member by  the one with largest $A$ {\em in the row immediately behind}; thus the car with largest $A$ in the row behind has advanced one row ahead; (iv) go on replacing a removed car by the one with largest $A$ in the row behind, until all the rows are exhausted (v) the car at the last row, which has gone one step ahead, will be replaced by a new member with $t_0=t$ and $A$ drawn from a uniform distribution in$(0,1)$. (The number of cars queued up again remains constant with time.) The numerical value of the exponent $\alpha$ is roughly the same as for the previous model (see Fig. 1(c, d)) with the difference that it is now rather insensitive to the width of the road. Thus, for $\sigma=2.0$ (0.5), the value of $\alpha$ ranges from 3.1 to 3.07 (1.7 to 1.55) as the width of the road varies from 10 to 50. 

It is natural to wonder what happens when more and more cars join the queue with time. Thus, at every time-step $t$, we add to the array, with a fixed probability $\rho$, another member $A_{L+1}$ with $N_{L+1} \in (0,1) $ so that the length of the array increases with time (linearly). The probability $P(\tau)$ now decays algebraically with $\tau$ only for small values of $\sigma$ and the dependence on the initial value of the number of cars is more pronounced. We must mention that when $\sigma=0$, that is, the aggressiveness of the drivers do not increase with waiting, the index $\alpha$ is 1.0 for $\rho=0$ and 1.5 for $\rho \ne 0$. This result was observed by Barabasi \cite{B1, B2}. \\

{\bf Connection between the exponents $\sigma$ and $\alpha$ : Analytic treatment :} First consider the case $\sigma=0$. Since the car with the highest value of $N$ is released at each time-step, the cars in the queue must have $N$ values distributed uniformly between zero and some limit $h(t)$ ($0 < h(t) < 1$) at a time $t$. Since the total number of cars introduced till now is $L+t$ and out of them $L$ are in the queue, we have $h(L+t) = L$. Now, the number of cars $dn(t_0,t)$ that were introduced during the interval $t_0$ to $t_0 + dt_0$ and released during the time interval $t$ to $t+dt$, is $L\,dh\, dt_0 = L^2/(L+t)^2\, dt \, dt_0$. Hence, the probability that the waiting time is $\tau=t-t_0$ is
\[ P(\tau) \propto \int_0^{\infty} \frac{L^2}{(L+\tau+t_0)^2} dt_0 = \frac{L^2}{L+\tau} \propto \frac{1}{\tau} \]
Thus, $\alpha=1$ for $\sigma=0$. For $\sigma >0$, the cars introduced in the interval $t_0$ to $t_0 + dt_0$ have $N$ values distributed uniformly between zero and $h$, where $h$ depends on both $t$ and $t_0$ and obeys the relation
\[ h(t - t_0)^{\sigma} = f(t) \]
where $f$ is some function to be determined. Since the total number of cars in the queue is $L$, we have
\[ L h(t_0=0,t) + \sum_{t_0=1}^{t_r-1} h(t_0,t) = L\]
For large $t$, the quantity $h(t_0=0,t)$ is small and can be ignored. In the case of $\sigma > 1$, the sum is convergent and reduces to $f(t)\zeta(\sigma)$ where $\zeta$ is the Riemann zeta function. This gives, $f = L/\zeta(\sigma)$ and 
\[ dn = \left(\frac{\sigma L}{\zeta(\sigma)} \right) \, \frac{1}{\tau^{\sigma + 1}} \,dt \]
which gives
\[ P(\tau) \propto \frac{1}{\tau^{\sigma + 1}} \]
It will be shown elsewhere that the relation $\alpha = \sigma + 1$ holds also for $0 < \sigma < 1$. One should note that in this arguement we have followed the mean field approach and have neglected all fluctuations. Hence it should be compared with the simulation results only for large $L$.\\

{\bf Discussion:} On the basis of the assumptions that (i) the movement of a vehicle at sub-normal speed occurs when it waits at a place where the road is narrow and only one of the many queued up cars can pass (ii) at a time from a queue, the car with the most aggressive driver passes through, and (iii) the aggresiveness is quantified by an attribute $A=N\tau^{\sigma}$ described in Eq. (\ref{defA}), we have shown (analytically and numerically) that the probability that a vehicle has to wait for a time $\tau$ is given by 
$P(\tau) \sim \tau^{-\alpha}$ as stated in Eq. (\ref{Ptau}) with $\alpha \approx \sigma+1$. The parameter $N$ randomly varies from 0 to 1 for different drivers and $\sigma$ is the same for all drivers. The empirical data for movement of taxis in two cities in India follows Eq. (\ref{Ptau}) with $\sigma=3.0$ to 3.5 and the same analysis for a city in Germany also follows the same pattern with $\sigma=1.5$. From this we conclude that the parameter $\sigma$ is a characteristic of a geographical region  and may signify some socio-cultural aspect of a region, as reflected in the driving pattern. More investigations are needed to test if our conclusions are true.\\
Two caveats : (1) Indeed, the empirical data may not always follow the algebraic pattern of Eq. (\ref{Ptau}). Our data for a city in India (namely, Mumbai) shows exponential decay of $P(\tau)$ rather than algebraic. (2) It may be possible that some hypotheses other than ours also reproduce the pattern of Eq. (\ref{Ptau}).\\

\begin{figure}
\subfigure{\includegraphics[width=5.0cm, angle=0]{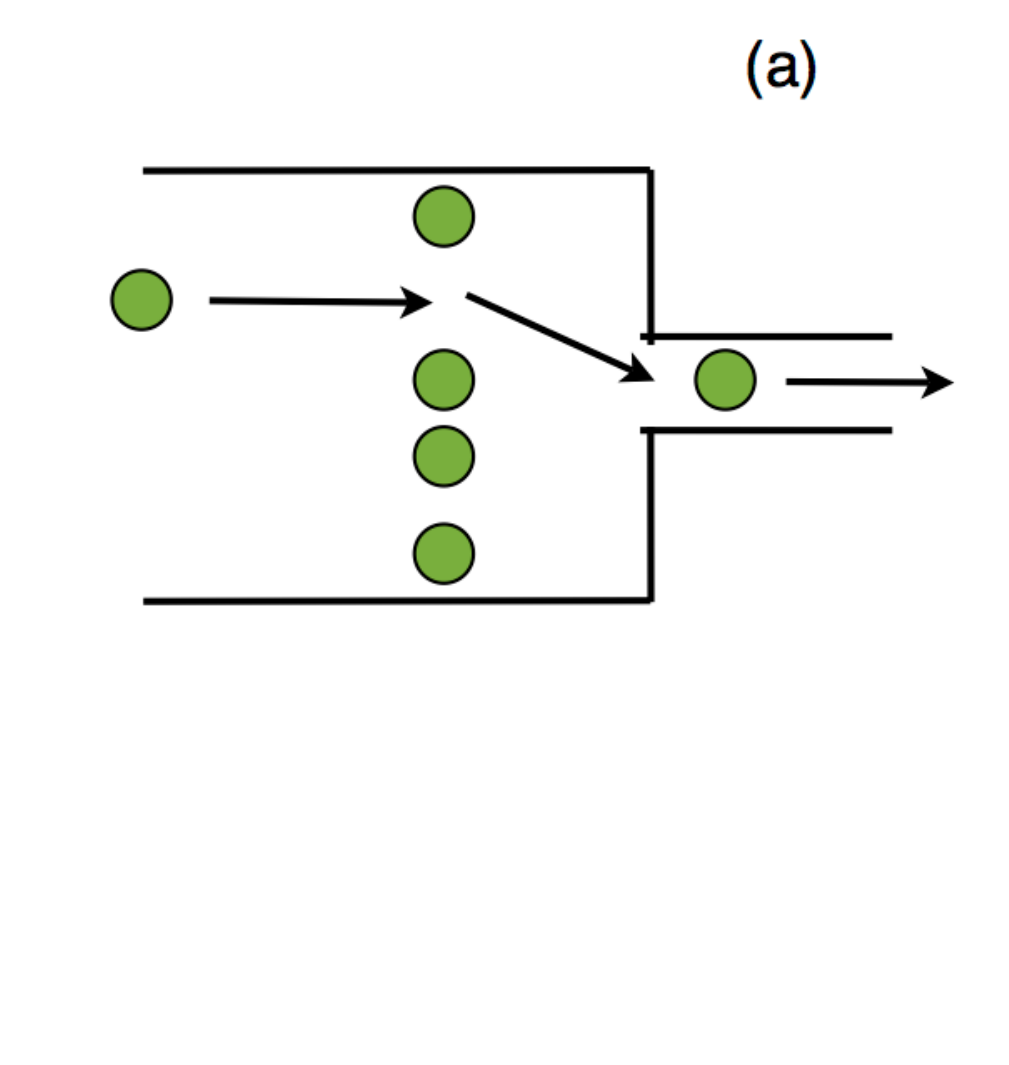}}
\subfigure{\includegraphics[width=9.0cm, angle=0]{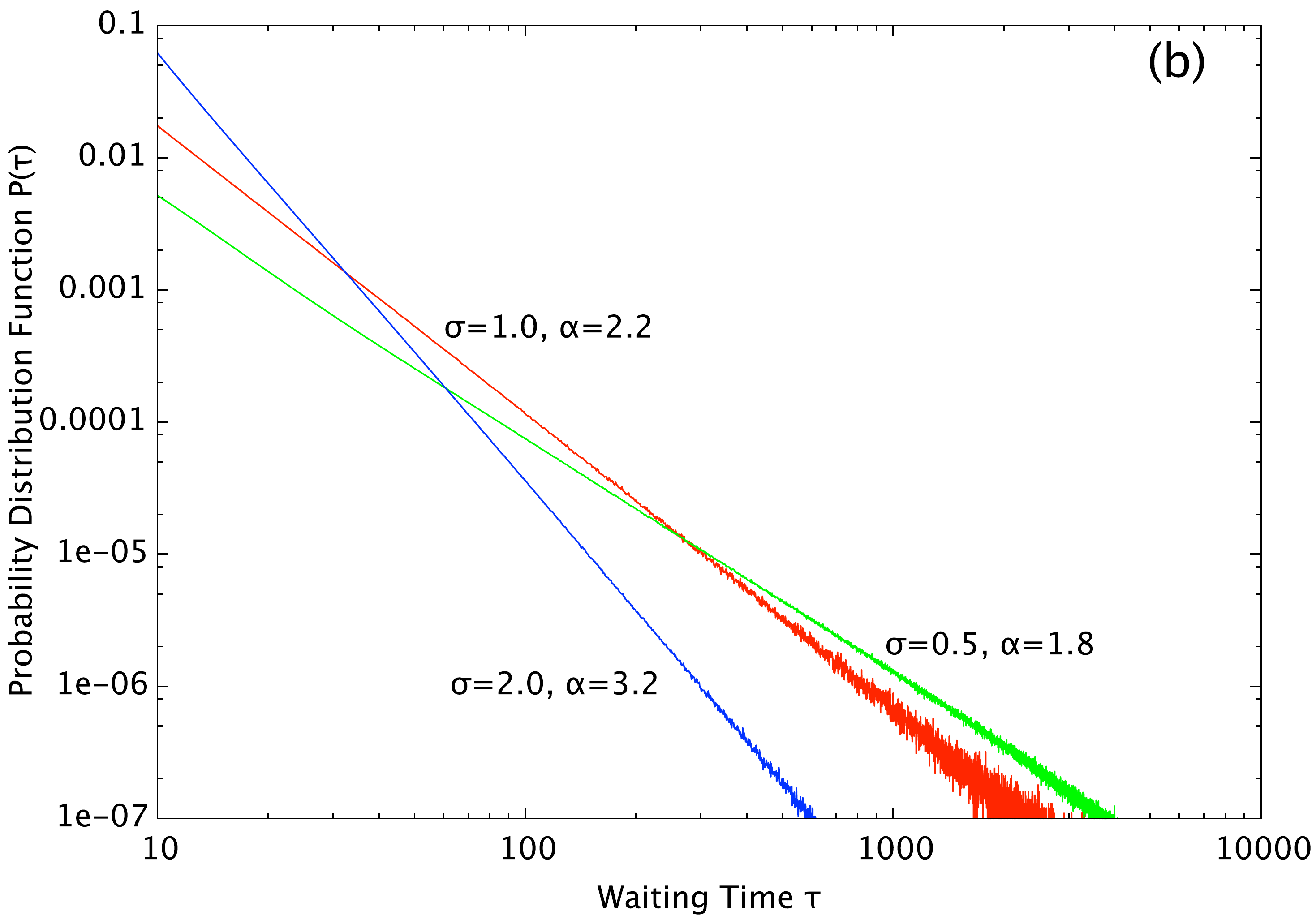}}
\subfigure{\includegraphics[width=5.0cm, angle=0]{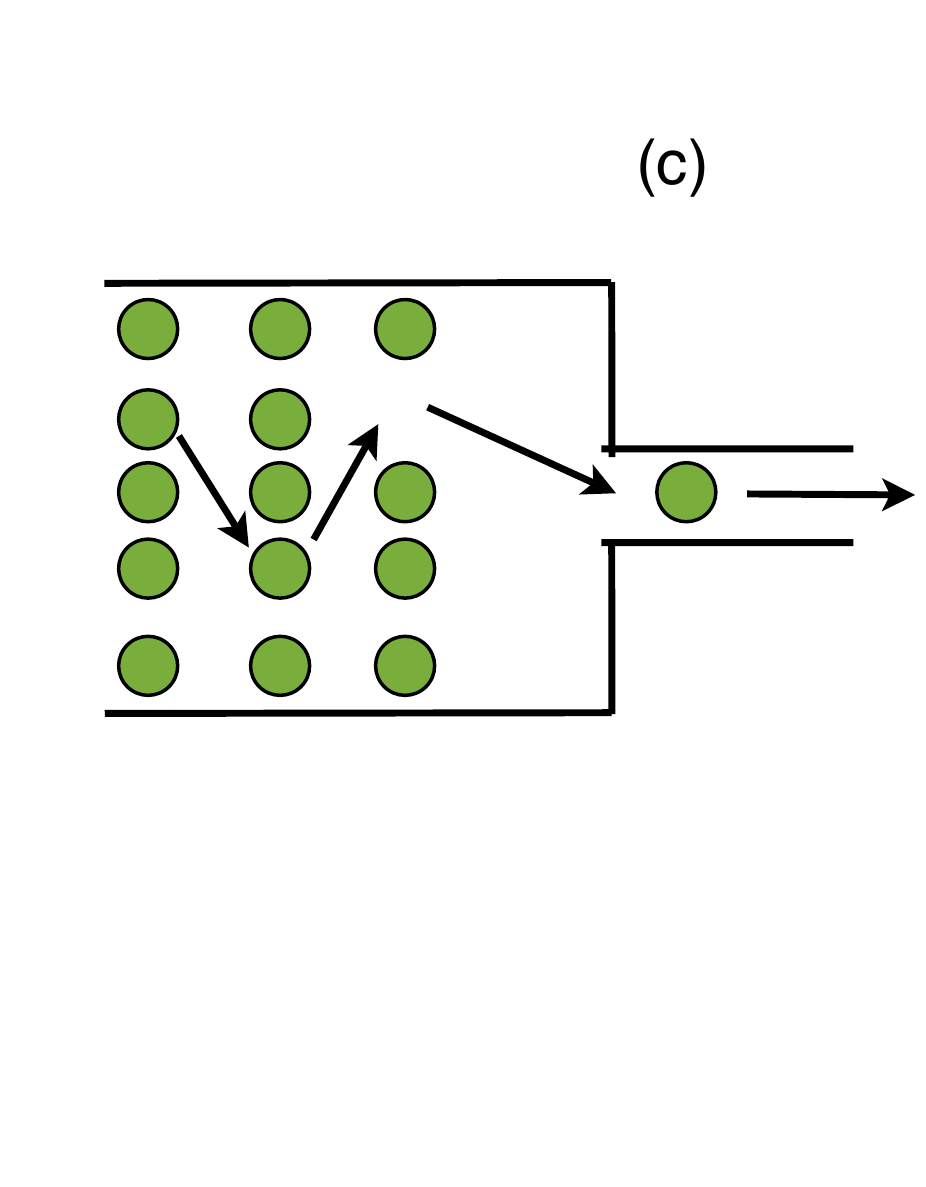}}
\subfigure{\includegraphics[width=9.0cm, angle=0]{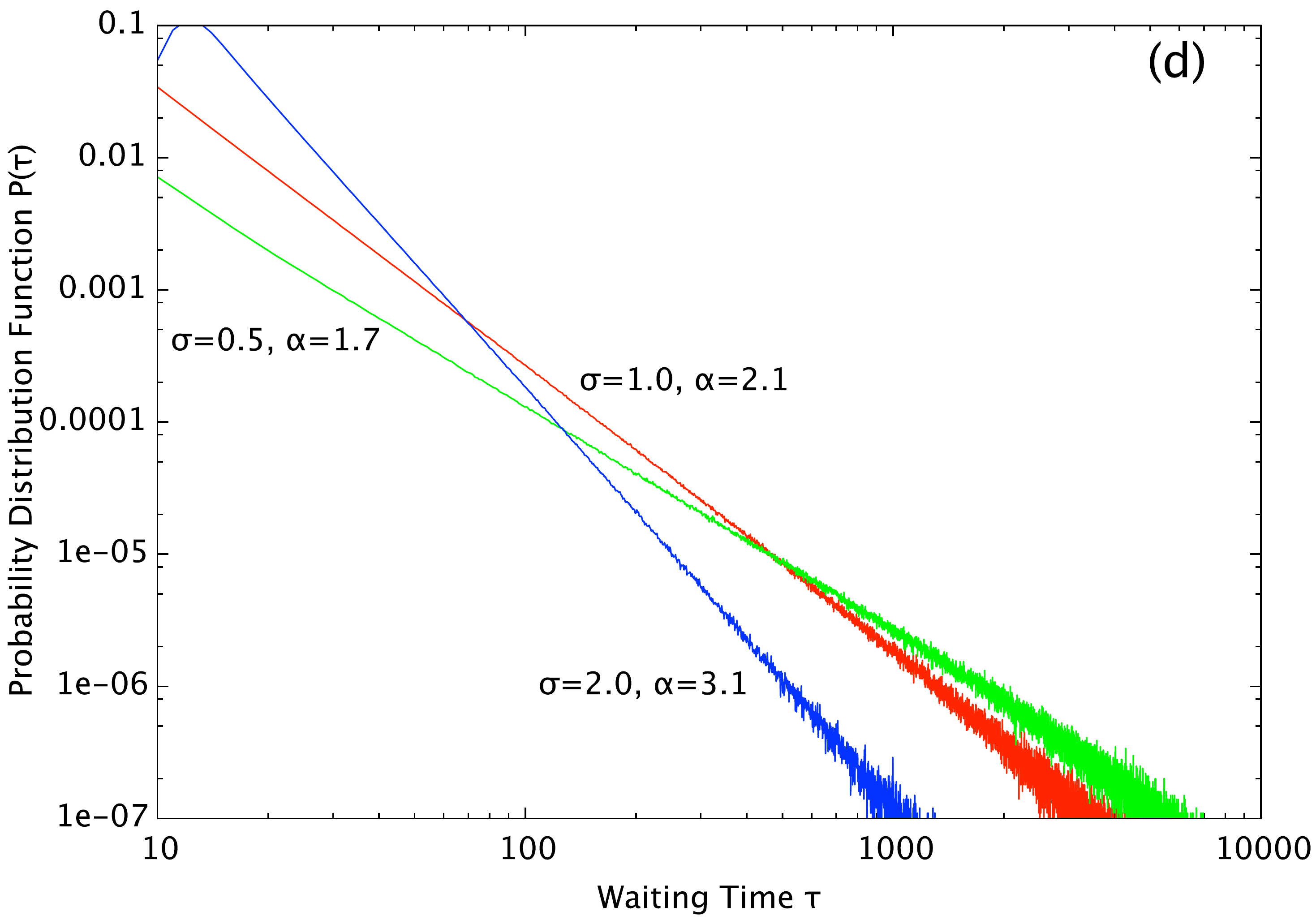}}
\caption{(a) The cars are accumulated before a constriction and the most aggressive one among them pass through at a time step. The removed car is immediately replaced by a new one. The aggressiveness increases following Eq. (\ref{defA}). (b) The waiting time distribution is algebraic with index $\alpha$. There are 10 cars at the constriction. The results were averaged over $10^5$ realisations. (c) Multiple rows of vehicles are queued up at the constriction. The most aggressive driver in the front row first passes through. This vacancy is filled up by the most aggressive car in the row behind. When the vacancy moves to the last row, it is filled up by a new car. (d) The waiting time distribution is algebraic. The simulation was done with two rows, each containing 10 cars but the results do not change if the number of rows is increased. The results were averaged over $10^4$ realisations.   }
\label{Fig-combined}
\end{figure}

\end{document}